\documentstyle[12pt]{article}

\begin{document}

\title{\bf\huge
A Suggestion for Modification of Vafa-Witten Theory
}

\author{
{\bf
Pei Wang}\\
\normalsize
Institute of Modern Physics,Northwest University\\
P.O.Box 105,Xian,710069,China
}

\maketitle

\begin{abstract}
Using the Mathai-Quillen formalism we reexamine the twisted
$N=4$ supersymmetric model of Vafa-Witten theory. Smooth out
the relation between the supersymmetric action and the path integral
representation of the Thom class.
\end{abstract}

\vskip 2truecm
{IMP-NWU-951201}
\newpage

\section{Introduction}
Recently Edward Witten made a breakthrough on the Donaldson theory by
his magical monopole equation [1]. He made also important developments
with N.Seiberg and C.Vafa [2,3,4] in his topological field theory[5].
These developments enrich the Donaldson theory enormously, so that
it can be explicitly described by using of quantum field theories in
variours supersymmetric models. In reference [4] the partition
functions of a twisted $N=4$ supersymmetric Yang-Mills
theory were computed on certain 4-manifolds and shown to satisfy
the remarkable Montonen-Olive "S-duality conjectures". One
important part of this story is that this model express the partition
function as an Euler characteristic of the moduli space of instantons.

However, it seems to us that two things might be improved. The first
is in their paper, the gauge transformation parameter, field
$\Phi $, is independent of the curvature derived from the connection
on a vector bundle with instanton moduli space (precisely, moduli space
dividing by gauge group) as its base space. We know that this connection
is defined through the action of gauge group[6,7]. There must
be close relation between them. The second is that the connection
we refer to is an object, which is hard to generalize to a superfield.
Of couse, there would be two ways to treat this object. One was to
consider it as a functional of a superfield. The other was thinking
of it as a component of a supermultiplets, which can be made of a new
superfield. But both lead to strange gauge transformation rules
and great complications. It is pleasure to smooth these problems.

In this paper, we would like to assume that the gauge parameter $\Phi $
is identified with the "curvature" (we will give the exact meaning of it
in the section 3) on the bundle with instanton base space. By using
of the Mathai-Quillen formula we develop the partition function of
Vafa-Witten theory in the Cartan model[8,9,10]. The result is equal
to a Yamron's action (with a missing term added)[11]. Under
suitable vanishing theorems, which were discussed carefully by Vafa
and Witten, this partition function is expressed by an Euler
characteristic (equivalently, a Thom class) strictly.

\section{Mathai-Quillen formulae of Thom class}
Since Witten and his colleagues have expressed a partition
function (more generally, correlation function) of topological
field as the Euler characteristic, it is quite nature that the
celebrated Mathai-Quillen formulae of Thom class[8] are very powerful
tools[10.12]. Let us recall something related. Let $E$ be an
oriented real vector bundle of rank $2m$, with base space $M$ and
standard fiber $V$, and
\begin{eqnarray}
s:M\rightarrow E\nonumber
\end{eqnarray}
\noindent be the zero section. The Euler class can be obtained through
the pull back of map $S$ from Thom class $\Phi (E)$,
\begin{eqnarray}
e(E)=s^*\Phi(E).\nonumber
\end{eqnarray}
\noindent Mathai and Quillen found an explicit formula for the universal
Thom class $U$ [8,10]. The criteria are 1) basic. 2) closed, 3) the
integration along the fiber is equal to unity, $\int _{V}U=1$. The
last criteron is connected with the definition of Thom class [13]:
the inverse of push forward of $1\in H^0(M)$ is defined as Thom class,
$\pi _*^{-1}=\Phi (E)\in H_{VRD}^{2m}(E)$, in which "VRD" denotes that
"rapid decrease along fiber", which is equivalent to the "vertical
compact supports". The formulae make use of equivariant cohomology in
which the Weil algebra plays the part of the universal bundle. In terms
of generators the Weil algbra $W(g)$ is expressed as follows[14]

\begin{eqnarray}
d_w\theta =\phi -\theta ^2, ~~ d_w \phi =\phi \theta -\theta \phi ,\\
\theta =T_i\theta ^i, ~~\phi =T_i\phi ^i, ~~T_i\in SO(2m).\nonumber
\end{eqnarray}
\noindent Through the Chern-Weil homomorphism (the basic form see for
example [8] or [10])
\begin{eqnarray}
\left (W(g)\otimes \Omega (V)\right ) _{basic}
\equiv \Omega _G(V)\rightarrow \Omega (E),\nonumber
\end{eqnarray}
\noindent we get the real differential form on the vector bundle
--partition function of the moduli space.

There are three expressions for the universal Thom class. The first
two are [8,10]
\begin{eqnarray}
U={\frac {1}{(2\pi )^m}}Pfaffian(\phi )e^{-{1\over {2\lambda }}
<\chi ,\chi >-\frac {1}{2\lambda }<\nabla \chi ,\phi ^{-1}\nabla \chi >}
\end{eqnarray}
\noindent and
\begin{eqnarray}
U=\frac {\lambda ^m}{(2\pi )^m}e^{-\frac {1}{2\lambda }<\chi, \chi >}
\int d\rho exp \left\{ \frac {1}{2\lambda }<\rho, \phi \rho >+
\frac {i}{\lambda }<\nabla \chi, \rho >\right\},
\end{eqnarray}
\noindent in which inner product is defined on $V$, $\rho $ is an
antighost variable, and
\begin{eqnarray}
\nabla \chi =dx+\theta \chi .
\end{eqnarray}
\noindent Following Vafa and Witten we also introduce a parameter
$\lambda $ in the formulae. From these two expressions criteria
1) and 3) are easily proved[8,10]. (The integration representation
(3) derives from (2) through a Gauss integration by use of
$\phi ^T =-\phi $ ). To check the criterion 2) so that we can write
down the partition function of the model later, we need the third
expression.

As usual we introduce again a zero ghost number field $\pi $ to
complete the enlarged equivariant cohomology complex
$W(g)\otimes \Omega (V)\otimes \Omega (\Pi V^*)$ (the $\Pi $
signifies that the coordinates are  to be regarded as anticommuting,
see [10]). But, instead of the common used deRham differential
$\delta \rho =\pi , ~\delta \pi =0$[10] we use the Vafa-Witten
mechanism. Let $\delta =d_w$ for Weil algebra, and
\begin{eqnarray}
\delta \rho =\pi -\theta \rho ,\nonumber \\
\delta \pi =-\theta \pi +\phi \rho .
\end{eqnarray}
\noindent It is easy to prove from (1) and (5) that $\delta ^2=0$. Let
$Q=d+\delta$, $d$ is the exterior differential on $\Omega (V)$, we
then have the third formula
\begin{eqnarray}
U=\frac {1}{(2\pi )^{2m}}\int _{V^*\times \Pi V^*}
\prod _{m=1}^{2m} d\pi _nd\rho _ne^{[Q,\Psi ]},
\end{eqnarray}
\noindent in which the gauge fermion is
\begin{eqnarray}
\Psi =\frac {1}{2\lambda }\rho (2i\chi -\pi ).
\end{eqnarray}
\noindent The close of Thom form $U$ follows from the following observation:

\begin{eqnarray}
(d+dw)\int (\cdots )=\int [Q, (\cdots )]=0.
\end{eqnarray}

When $U$ is transferred to a partition function of a real field model
through the Chern-Weil homomorphism, the generators of Weil algebra
will correspond the connection $\Theta $ and curvature $\Phi $ on the
vector bundle of moduli space respectively:
\begin{eqnarray}
\theta \rightarrow \Theta,~~\phi \rightarrow \Phi .
\end{eqnarray}

\section{Gauge invariance and Cartan model}
Now, let us consider the case $\delta ^2\not =0$, corresponding a gauge
transformation. Let $\cal {U}$ denotes the priciple bundle consisting of
Yang-Mills potentials. Locally it can be parametrized by
\begin{eqnarray}
A=g^{-1}ag+g^{-1}dg,~~a\in orbit{\cal U}/{\cal G},~~
g\in {\rm gauge~ group}{\cal {G}}.
\end{eqnarray}
\noindent It gives $(\delta d+d\delta =0)$
\begin{eqnarray}
\delta A=g^{-1}\delta ag-D_{A}\left ( g^{-1}\delta g\right ),
\end{eqnarray}
\noindent where
\begin{eqnarray}
D_A\Lambda =d\Lambda +A\Lambda -(-)^p\Lambda A,
\end{eqnarray}
$p$ is the parity of $\Lambda $(counting of both $d$ and $\delta $).
A connection can be introduced on $\cal U$ [6].
\begin{eqnarray}
\Theta =-G_AD_A^{\dagger }\delta A,~~G_A=(D_A^{\dagger }D_A)^{-1},
\end{eqnarray}
\noindent such that
\begin{eqnarray}
\Theta |_{fiber}=g^{-1}\delta g.
\end{eqnarray}
\noindent Up to a zero mode of $D_A^{\dagger }$ we may write
\begin{eqnarray}
\delta A=-D_A\Theta .
\end{eqnarray}
\noindent From this connection a formal "curvature" is derived as
\begin{eqnarray}
\Phi =\delta \Theta +\Theta ^2,~~ \Phi |_{fiber}=g^{-1}
\delta ^2g.
\end{eqnarray}
\noindent We can define the local gauge transformation in case at hand,
\begin{eqnarray}
\delta _G=\epsilon \delta ^2
\end{eqnarray}
\noindent wher $\epsilon $ is a Grassmann number anticommuting with $d$
and $\delta $, so that
\begin{eqnarray}
\delta _GA=g^{-1}\delta _Gag-D_A(g^{-1}\delta _Gg).
\end{eqnarray}
\noindent And eq.(15) leads to
\begin{eqnarray}
\delta _GA=-D_A\epsilon \Phi =\epsilon D_A\Phi .
\end{eqnarray}
\noindent However, this $\Phi $ does not transform like a curvature under
a "bundle gauge transformation" $\Theta \rightarrow \Theta '=
h^{-1}\Theta h+h^{-1}\delta h$. So we
have to define a real curvature
$\Omega $ as follows. Let $\Theta =\delta A^i\Theta _i$, then we have
\begin{eqnarray}
\Phi =\delta ^2A^i\Theta _i +\Omega,~~\Omega={1\over 2}\delta A^i
\delta A^j\Omega _{ij}.
\end{eqnarray}
\noindent Any way we may have
\begin{eqnarray}
\delta A=\Psi , ~~\delta \Psi =D_A\Phi .
\end{eqnarray}

If we again use the correspondence between Weil algebra and connection
as well as curvature (9), the Weil algebra must change to
\begin{eqnarray}
d_w'\theta =\phi -\theta ^2,~~d_w'={d_w'} ^2\theta ^2+\phi \theta -
\theta \phi .
\end{eqnarray}
\noindent Changing eq.(5) also to (see [4])
\begin{eqnarray}
\delta \rho =\pi -\theta \rho ,\nonumber \\
\delta \pi =-\theta \pi +\Omega \rho ,
\end{eqnarray}
\noindent we find that
\begin{eqnarray}
[Q,\Psi ]=-\frac {1}{2\lambda }(\pi -i\chi )^2
-\frac {1}{2\lambda }<\chi ,\chi >+\frac {1}{2\lambda }
<\rho ,\Omega \rho >+\frac {i}{\lambda }<\nabla \chi ,\rho >.
\end{eqnarray}
\noindent So we have the same expressions in the first two formulae for Thom
class (2) and (3), with $\Omega $ instead of $\Phi $ only. And the
criteria 1) and 3) for Thom form are satisfied. To require that the form
is closed, the gauge fermion $\Psi $ must be gauge invariant, i.e.
$\left\{ Q[Q,\Psi ]\right\} =0$. As a matter of fact, however, we find that
\begin{eqnarray}
\left\{ Q[Q,\Psi ]\right\} =-\frac {1}{2\lambda }
\left\{ 2i\rho (\Omega -\Phi )\chi +\delta ^2A^i
\delta A^j)\rho \Omega _{ij}\rho \right\} \not = 0.\nonumber
\end{eqnarray}
\noindent To overcome this difficult we have to choose the Cartan
model $S(g^*)\otimes \Omega (V)$,
\begin{eqnarray}
\theta \rightarrow 0,~~d\rightarrow d_c=d-i_{\phi }.
\end{eqnarray}
\noindent where $i_{\phi }$ is the contraction operator [8,9,10]. In
the case at hand, $\Omega $ is undistinguishable from $\Phi $. Let
$\delta =d_c$ on $\Omega (V)$, and
\begin{eqnarray}
\delta \rho =\pi ,\nonumber \\
\delta \pi =\phi \rho .
\end{eqnarray}
\noindent It is easy to check that
\begin{eqnarray}
\delta \Psi =[Q_c,\Psi ]=-\frac {1}{2\lambda }
(\pi -i\chi )^2-\frac {1}{2\lambda }<\chi ,\chi >
+\frac {1}{2\lambda }<\rho ,\phi \rho >+
\frac {i}{\lambda }<d\chi ,\rho >.
\end{eqnarray}
\noindent Keeping correct the properties 1) and 3), it is also clear that
\begin{eqnarray}
\left\{ Q_c[Q_c,\Psi ]\right\} =0,
\end{eqnarray}
\noindent so that
\begin{eqnarray}
U=\frac {1}{(2\pi )^m}\int _{V^*\times \Pi V^*}
\prod _{n=1}^{2m}d\rho _nd\pi _ne^{[Q_c,\Psi ]}
\end{eqnarray}

\noindent is closed. Cartan model is also a favorable matter, because of
that to construct Donaldson invariants from quantum field theory
we need [5,2,12]
\begin{eqnarray}
\delta \phi =0.
\end{eqnarray}

\section{Projection by the gauge freedom}
When we treat the topological Yang-Mills theory, the gauge invariance
must be considered. The moduli space should be divided by
$G=G_{gauge}$,
and in fact we will compute the Euler characteristic of the quotient
bundle $E/G$. This can be done by introducing two new fields
$\bar {\phi },\eta $ of ghost number -2 and -1 respectively. And let
\begin{eqnarray}
\delta {\bar {\phi }}=[Q_c,{\bar {\phi }}]=\eta ,\nonumber \\
\delta \eta =[Q_c, \eta ]=[{\bar {\phi}},\phi ].
\end{eqnarray}
\noindent The consequence of the gauge invariance leads to a projection
gauge fermion[10]
\begin{eqnarray}
\Psi _{proj}=\psi D\bar {\phi },
\end{eqnarray}
\noindent where $\psi $ is defined in eq.(21). The Thom class of interest
becomes the following form ($t=dimG$)
\begin{eqnarray}
\Phi (E/G)&=&\frac {1}{(2\pi )^{m+t}i^t}\int \prod
d\bar {\phi }d\eta d\rho d\pi e^{[Q_c,\Psi _{tot}]},\\
\Psi _{tot}&=&\Psi _{local}+\Psi _{proj.}+\Psi _{non.}
\end{eqnarray}
\noindent in which $\Psi _{local}=\Psi $ is the localized gauge
fermion (7) obtained earlier, and $\Psi _{non.}$ is a possible
nonminimal term[4].

\section{$N=4$ twisted supersymmetric Yang-Mills model}
Now we are in the position to construct the concret model. Let
$\cal {M}$ denote the moduli space of antiselfdual instantons,
$\cal {V}$
denote the fiber, what we are interested in is the vector bundle
$\varepsilon =\cal {M}\times _GV$. Through out the Chern-Weil
homomorphism we will introduce various fields to realize the Thom
form, the partition function. Before that, as is pointed out
by Vafa and Witten, when the parameter $\lambda $ approaches to
zero the integral leads to a trivial results but with signs, a standard
theorem that the Euler class of a bundle is computed by counting the
zeros of a section, weighted with signs. To avoid the cancellation
of Euler class due to different signs they invent a way
of doubling the number of fields.

Therefore we have all of the twisted $N=4$ supersymmetric fields[4]
(The ghost numbers are marked in the brackets):

\begin{eqnarray}
A_i,~B_{ij}~(0)& &{\rm coordinates~of~moduli~space;}\nonumber \\
\psi _i,~\tilde {\psi }_{ij}~(1)& &{\rm tangent~to~the~moduli~manifold,}
\nonumber\\ & &{\rm the~super~partners~of~A_i,~B_{ij}~respectively};
\nonumber \\
\chi _{ij},~\tilde {\chi }_i~~(-1)& &{\rm for~\rho,~coordinates~of~fiber};
\nonumber \\
H_{ij},~\tilde {H}_i~~(0)& &{\rm for~\pi ,~auxiliary~fields,~the~super}
\nonumber \\
& & {\rm partners~of~\chi _{ij},~\tilde {\chi }_i~respectively;}
\nonumber \\
\phi (2),\tilde {\phi }(-2),\eta (-1)& &{\rm associated~with~the
{}~symmetry~group}~G,\nonumber \\
& &{\rm the~same~fields~we~have~used};\nonumber \\
C(0), \xi (1) & &{\rm associated~with~the~symmetries,}\nonumber \\
& &{\rm reprsenting~the~horizontal~projection,}\nonumber
\end{eqnarray}

\noindent in which the supermultiplet ($A_i,\psi _i $) have presented
in eq.(21). Moreover, we have also to substitute the coordinates $\chi $
in the Thom form by the section pairs[4]
\begin{eqnarray}
S_{ij}=F_{ij}^{\dagger }+\frac {i}{2} [C,B_{ij}^{\dagger }]
+\alpha [B_{ki}^{\dagger }, B_j^{\dagger k}],
\end{eqnarray}
\noindent and
\begin{eqnarray}
k_i=2D^jB_{ij}^{\dagger }+{1\over {2i}}D_iC,
\end{eqnarray}
\noindent in which $\alpha $ is a constant, $F_{ij}^{\dagger }$
and $B_{ij}^{\dagger }$ are antiself-dual part of $F_{ij}=
\partial _iA_j-\partial _jA_i+[A_i,A_j]$ and $B_{ij}$( we will
omit the symbol $\dagger $ for $B_{ij}^{\dagger }$ for
simplicity). Noting that the projection gauge fermion needs also
doubled we obtain the partition function of the model as
follows
\begin{eqnarray}
Z=N\int [d\Sigma ]exp
\left\{ Q_c(\Psi _{local}+\Psi _{proj.}+\Psi _{non.}\right\},
\end{eqnarray}
\noindent where $[d\Sigma ]$ is the measure consisting of all of
the thirteen fields, $N$ is the normalization constant,
\begin{eqnarray}
\Psi _{local}&=&T_r\left\{\chi ^{ij}
(2iF_{ij}^{\dagger }-[C,B_{ij}]+2i\alpha [B_{ki},{B_j}^k]-H_{ij}\right.
\nonumber \\
& &\left. +\tilde {\chi }^i(4iD^jB_{ij}+D_iC-\tilde {H}_i)\right\},
\end{eqnarray}
\noindent and
\begin{eqnarray}
\Psi _{proj}=-T_r \left\{ \psi ^iD_i\tilde {\phi }+
\tilde {\psi }^{ij}[B_{ij},\tilde {\phi }]\right\},
\end{eqnarray}
\noindent one may also have
\begin{eqnarray}
\Psi _{non}=\beta T_r\left\{ \eta [\tilde {\phi },\phi ]
+2C[\tilde {\phi }, \xi ]\right\}, ~\beta =const.
\end{eqnarray}
\noindent It is clear that if there are suitable vanishing theorems
as was discussed carefully by Vafa and Witten,i.e.
$B_{ij}=C=0$ [4], set $\beta =0$ the partition function identifies
a Thom class (equivalently, an Euler characteristic).

\section{BRST and anti BRST transformations}
One more point we would like to explain is that the twisted $N=4$
system has $N=2$ topological symmetry and an $SU(2)$ symmetry.
We have not only a $Q_1\equiv Q_c$, but $Q_A~~A=1,2$ pair. One is
thought of a BRST operator, the other can be identified with an
anti BRST charge[11]. Let us introduce following superfields:
\begin{eqnarray}
T^i=\partial ^i+A^i+\theta ^A\psi _A^i+
\frac {{\epsilon }_{AB}\theta ^A\theta ^B}{2}\tilde {H}^i,
\end{eqnarray}
\noindent in which $\psi _1^i=\psi ^i, \psi _2^i=\tilde {\chi }^i$;
\begin{eqnarray}
Y^{ij}=B^{ij}+\theta ^A\chi _A^{ij}+
\frac {\epsilon _{AB}\theta ^A\theta ^B}{2}H^{ij},
\end{eqnarray}
\noindent in which $\chi _1^{ij}=\tilde {\psi }^{ij},
\chi _2^{ij}=\chi ^{ij};$
\begin{eqnarray}
X_{AB}=\phi _{AB}+{1\over 2}(\theta _A\eta _B
+\theta _B\eta _A)-
\frac {\epsilon _{CD}\theta ^C\theta ^D}{4}[\phi _{AE},{\phi ^E}_B],
\end{eqnarray}
\noindent in which $\phi _{11}=\phi ^{22}=\phi $,
$\phi _{22}=\phi ^{11}=\tilde {\phi }$, and $\phi _{12}
=\phi _{21}=-\phi ^{12}=-\phi ^{21}=C.$ The supersymmetry
transformations are determined by
\begin{eqnarray}
Q_A={\frac {\partial }{\partial \theta ^A}}+\theta ^B
[\phi _{AB},\cdot ]+\frac {\epsilon _{BC}
\theta ^B\theta ^C}{4}[\eta _A,\cdot ],
\end{eqnarray}

\noindent in which we add a two $\theta $ term so that the three
transformation rules derived from the superfield $X_{AB}$ are
selfconsistant. And it also gives the same transformation laws as
Yamron's[11]. In fact, the transformation laws may write as
\begin{eqnarray}
\delta S=\epsilon ^A[Q_A, S],~~S=T^i,Y^{ij},X_{AB}.
\end{eqnarray}
\noindent More detail we get

\begin{eqnarray}
\delta A^i&=&\epsilon ^1\psi ^i+\epsilon ^2\tilde {\chi }^i,\nonumber \\
\delta \psi ^i &=&\epsilon ^1D^i\phi
-\epsilon ^2(\tilde {H}^i-D^iC),\nonumber \\
\delta {\tilde {\chi }}^i&=&\epsilon ^1
(\tilde {H}^i+D^iC)+\epsilon ^2D^i\tilde {\phi },\nonumber \\
\delta {\tilde {H}}^i&=&\epsilon ^1
\left ( [C,\psi ^i]-[\phi ,\tilde {\chi }^i]
-{1\over 2}D^i\xi \right )+\epsilon ^2\left (
[{\tilde {\phi }},\psi ^i]-[C,\tilde {\chi }^i]
-{1\over 2}D^i\eta \right ),\nonumber \\
\delta B^{ij}&=&\epsilon ^1\tilde {\psi }^{ij}+\epsilon ^2\chi ^{ij},
\nonumber \\
\delta {\tilde {\psi }}^{ij}&=&\epsilon ^1
[B^{ij},\phi ]-\epsilon ^2\left (H^{ij}+[C,B^{ij}]\right ),\nonumber \\
\delta \chi ^{ij}&=&\epsilon ^1\left (H^{ij}-
[C,B^{ij}]\right )-\epsilon ^2[\tilde {\phi },B^{ij}],\nonumber \\
\delta H^{ij}&=&\epsilon ^1
\left ( [C,\tilde {\psi }^{ij}] -[\phi ,\chi ^{ij}]
+{1\over 2}[\xi ,B^{ij}]\right )\nonumber \\
& &+\epsilon ^2\left ([{\tilde {\phi }},\tilde {\psi }^{ij}]
-[C,\chi ^{ij}] +{1\over 2}[\eta ,B^{ij}]\right ),\nonumber \\
\delta \phi &=&-\epsilon ^2\xi ,\nonumber \\
\delta C&=&{1\over 2}(\epsilon ^1\xi -\epsilon ^2\eta ),\nonumber \\
\delta \tilde {\phi }&=&\epsilon ^1\eta ,\nonumber \\
\delta \eta &=&\epsilon ^1[\tilde {\phi },\phi ]
+2\epsilon ^2[\tilde {\phi },C],\nonumber \\
\delta \xi &=&2\epsilon ^1[C,\phi ]
+\epsilon ^2[\tilde {\phi },\phi ].
\end{eqnarray}

\noindent For the parts proportional to $\epsilon ^1$ we find
the transformation laws we used before. Using the anti-BRST
transformations (proportional to $\epsilon ^2$ ) we
can prove that
\begin{eqnarray}
\Psi _{tot}&=&Q_2W,\\
W&=&T_r\left\{ B^{ij}
(2iF_{ij}-H_{ij})+{\frac {2i\alpha }{3}}B_{ij}
[{B_k}^i, B^{jk}]\right.\nonumber \\
& &\left.+\psi ^i\tilde {\chi }_i+\beta \phi _{AB}
[{\phi _C} ^A,\phi ^{BC}]\right\} .
\end{eqnarray}

In conclusion, we have
\begin{eqnarray}
Z=N\int [d\Sigma ]e^{{1\over 2}\left\{
Q^A[Q_A,W]\right\} }.
\end{eqnarray}
\noindent It agrees with the action obtained by Yamron except an added
term $B_{ij}[{B_k}^i,B^{jk}]$.

\section{Summary}
To sum up, we have derived the partition function of the twisted $N=4$
supersymmetric Yang-Mills theory, and it reduces to the Euler
characteristic under suitable vanishing theorems. We can also integrate
over auxiliary fields, the main part of the result have already
presented in Vafa-Witten's paper[4], i.e. $|s|^2+|k|^2$. One
interesting thing is to find the corresponding formulae in the
Weil model. This relats to that how to combine a basic form
with the gauge transformation, or how to find a gauge invariant
gauge fermion. We have not found the answer yet.

\newpage

\noindent {\large \bf Acknowlegements}

The author would like to thank Ke Wu for hospitality at Institute
of Theoretic Physics in Beijing, and benefit discussions, especially
drawing his interest to the Mathai-Quillen formalism. He also
would like to thank Yi-hong Gao for valuable introduction talk
about Donaldson-Witten theory and bringing him the preprint[10]. The
work is supported in part by the National Science Fund of China.

\end{document}